\documentclass[10pt,twocolumn,superscriptaddress,showpacs,pra,aps]{revtex4-1}

\usepackage{amsmath}    
\usepackage{graphicx}   
\usepackage{verbatim}   
\usepackage{color}      
\usepackage{subfigure}  
\usepackage{hyperref}   
\usepackage{units}

\begin{document}

\title{Efficient demagnetization cooling of atoms and its limits}

\author{Valentin V. Volchkov} 
\affiliation{5. Physikalisches Institut, Universit\"at Stuttgart, Pfaffenwaldring 57 D-70550 Stuttgart, Germany}
\affiliation{Laboratoire Charles Fabry, Institut d'Optique, CNRS, 2 Avenue Augustin Fresnel 91127 Palaiseau cedex, France}
\author{Jahn R\"uhrig} 
\affiliation{5. Physikalisches Institut, Universit\"at Stuttgart, Pfaffenwaldring 57 D-70550 Stuttgart, Germany}
\author{Tilman Pfau}
\affiliation{5. Physikalisches Institut, Universit\"at Stuttgart, Pfaffenwaldring 57 D-70550 Stuttgart, Germany}
\author{Axel Griesmaier}
\affiliation{5. Physikalisches Institut, Universit\"at Stuttgart, Pfaffenwaldring 57 D-70550 Stuttgart, Germany}

\date{\today}


\date{\today}

\begin{abstract}
Demagnetization cooling relies on spin-orbit coupling that brings motional and spin degrees of freedom into thermal equilibrium. 
In the case of a gas, one has the advantage that the spin degree of freedom can be cooled very efficiently using optical pumping. 
We investigate demagnetization cooling of a chromium gas in a deep optical dipole trap over a large temperature range and reach high densities up to $5\times 10^{19} m^{-3}$. 
We study the loss mechanism under such extreme conditions and identify excited-state collisions as the main limiting process. 
We discuss that in some systems demagnetization cooling has a realistic potential of reaching degeneracy by optical cooling only. 
\end{abstract}

\maketitle
\begin{figure}
	\includegraphics[scale=0.8]{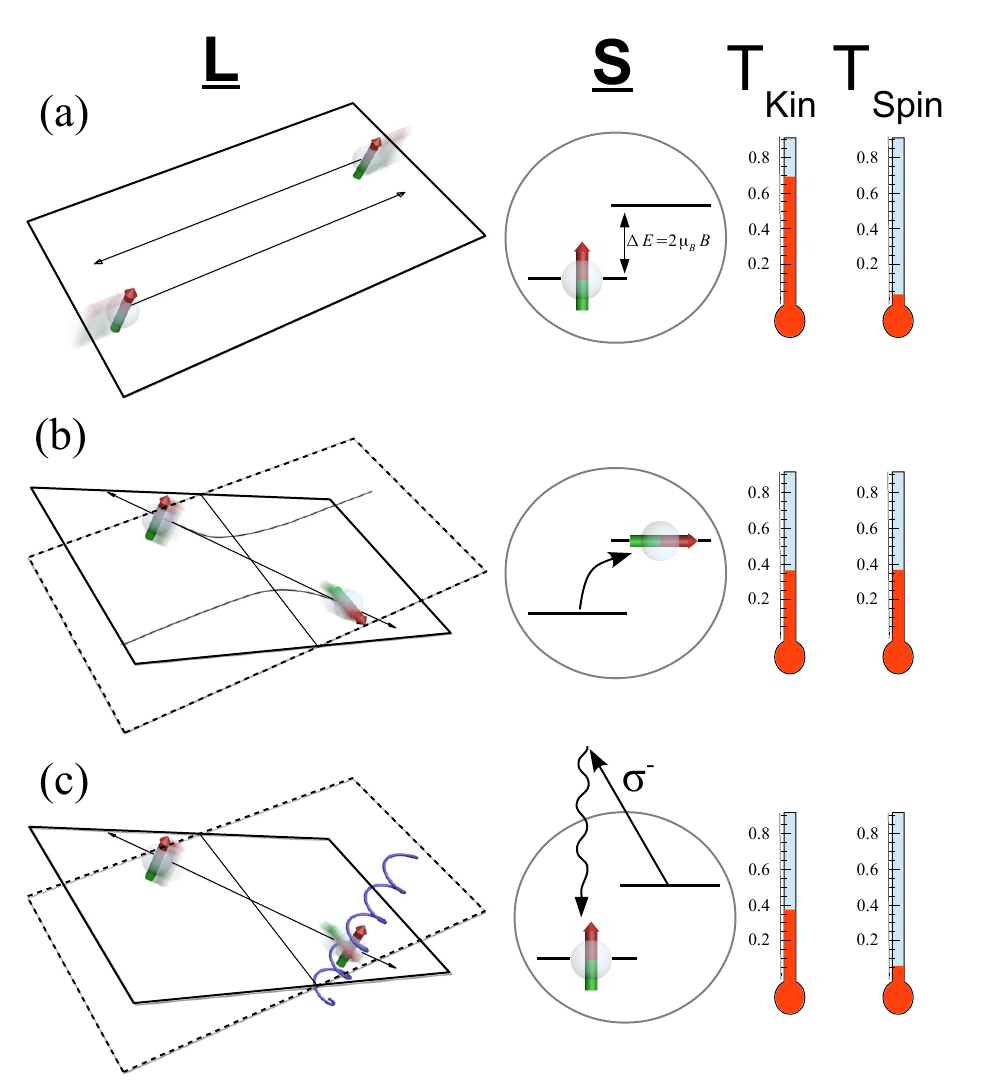}
	\caption{\label{fig:cartoon} Illustration of the demagnetization cooling. 
	Two atoms in a spin-polarized state approach each other with finite angular momentum \underline{\textbf{L}} (a). 
During their (inelastic) collision the orbit is coupled to the spin \underline{\textbf{S}} via dipole-dipole interaction. 
The plane of the outgoing atoms in panel (b) is inclined with respect to the plane of the incoming atoms, which indicates the non-conservation of the orbital angular momentum. 
Kinetic energy is converted into Zeeman energy leading to thermalization of the motional and spin degrees of freedom, as shown in panel (b). 
Finally, optical pumping cools the spin degree of freedom and dissipates the energy by the scattered photon (c). }
\end{figure}

In recent years, Bose-Einstein condensation of non-alkali-metal elements has attracted growing interest because of their magnetic and electronic properties \cite{Lu11b,*Aikawa12,*Stellmer09,*Kraft09,*Takasu03,*griesmaier05}. 
On the one hand, the very magnetic species chromium, erbium, and dysprosium allow extending the possible studies of dipolar interacting and  strongly correlated quantum matter \cite{Goral02,*Pedri05,*Santos03}. On the other hand, the complex electronic structures of non-alkali-metal elements gives rise to new laser cooling methods \cite{Katori99,Frisch12}. 
Renewed interest in optical cooling methods originates from progress in laser cooling of molecules \cite{zeppenfeld12,shuman10} and the possibility of using narrow linewidth transitions for standard laser cooling techniques \cite{mcclelland06}. 
The motivations are manifold. 
On the one hand, optical cooling may pave the way to degeneracy of ground-state molecules - a goal pursued by many groups. 
On the other hand, new \textit{lossless} laser cooling methods may be considered as an alternative to evaporative cooling, allowing one to boost the number of atoms of a degenerate gas. 
Finally, reaching Bose-Einstein condensation by optical means only\cite{BAR96,santos01,Stellmer13} is a long-standing goal in the field.
In this context, we investigate demagnetization cooling of cold gases, a technique combining dipolar collisions and optical pumping \cite{kastler50}. 
The cooling of the internal degree of freedom (spin) is realized by optical pumping, while dipolar relaxation provides the mechanism for rethermalization of the internal and the external degrees of freedom.  
Starting from a polarized sample in the lowest magnetic substate [Fig. \ref{fig:cartoon}(a)], the kinetic energy of a pair of inelastically colliding atoms is reduced by $\Delta{}E$ for each particle that is promoted to the higher magnetic substate [Fig. \ref{fig:cartoon}(b)].
The Zeeman energy $\Delta{}E=2\mu_BB$, where $B$ is the applied magnetic field, corresponds to the energy splitting between neighboring magnetic substates. 
Thus, the dynamics of the inelastic collisions is governed by the magnitude of the magnetic field, since only atoms with sufficient relative kinetic energy may undergo a spin flip. 
Therefore, the energy splitting $\Delta{}E$ is typically on the order of the temperature of the sample. 
Optical pumping back to the lowest substate cools the spin degree of freedom and dissipates the Zeeman energy [Fig. \ref{fig:cartoon}(c)].
Furthermore, the optical pumping transition and the polarization of the pumping light are chosen such that the atoms in the internal magnetic ground-state do not couple to the light.  
In a proof-of-principle experiment demagnetization cooling of a chromium gas has been demonstrated by our group \cite{fattori06}. 
A moderate cooling effect was observed in this experiment; however, the limiting effect remained unidentified.	
According to the theory \cite{hensler05} temperature on the order of the recoil temperature  ($\mathrm{T_R}$) should be attainable. 
This raises the question of whether quantum degeneracy can be reached using demagnetization cooling only. 
\par
In this paper, we present a quantitative study of demagnetization cooling of $^{52}$Cr in a temperature range from \unit[90] to \unit[6]{$\mu K$}. 
The cooling rate we have achieved has improved by more than 1 order of magnitude compared to previous work. 
We also identify light-assisted losses as the dominating limiting mechanism for this regime. 
Finally, we discuss the more favorable conditions for demagnetization cooling of dysprosium (or any other highly magnetic lanthanide).
\par
We have performed the experiments in a deep optical dipole trap (ODT), created by \unit[90]{W} of a \unit[1070]{-nm} laser beam, focused to \unit[30]{$\mu$m}. 
The resulting trap depth corresponded to \unit[1.3]{mK}, with trapping frequencies  $\omega_x=\omega_y=2\pi\times$\unit[5.5]{kHz} and $\omega_z=2\pi\times$\unit[40]{Hz}. 
We loaded the ODT from a guided atomic beam as described in Refs. \cite{falkenau11,volchkov13}. 
We started the demagnetization cooling sequence with $N_0=8\times10^5$ atoms at a temperature of $T_0=$\unit[90]{$\mu$K} by applying a homogeneous magnetic offset field in the  $x$ direction and shining in optical pumping light aligned along the axis of the magnetic field. 
The magnitude of the offset field was reduced stepwise during the cooling sequence in order to adjust the Zeeman splitting to the decreasing thermal energy. 
We used magnetic field coils in all directions to compensate for external static fields and to precisely overlap the axis of the offset field with the propagation axis of the pumping light. 
We created circularly polarized light with $I_{\sigma^-}/I_{\sigma^+}>4000$ using a combination of three retardation plates ($\lambda/4$ - $\lambda/2$ - $\lambda/4$) compensating for possible imperfections of the polarization optics (polarizing cube, retardation plates) as well as for minor birefringence of the entry view port. 
As a consequence of the polarization, only $\sigma^-$ transitions were driven during the optical pumping process. 
This ensures that the lowest magnetic substate with $m_J=-3$ is a dark state and does not scatter any incoming optical pumping light. 
The optical pumping was implemented using the $\mathrm{^7S_3\rightarrow{}^7P_3}$ transition at \unit[427.6]{nm}. 
After a variable cooling time we released the atoms from the trap and took an absorption image. 
From this image we determined the number of atoms and the temperature of the sample.
\par
\begin{figure}
	\includegraphics[scale=1]{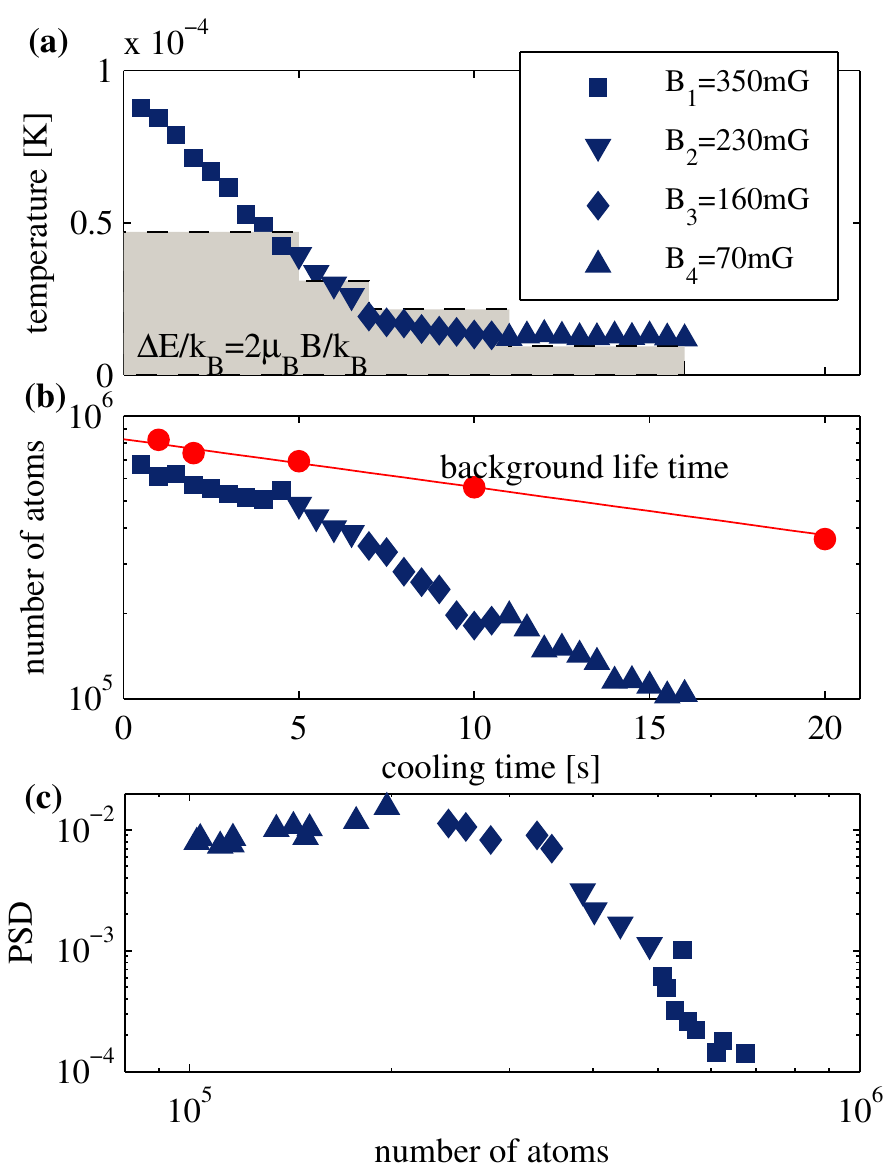}
	\caption{\label{fig:coolingplot} (a) Time evolution of the temperature. 
	The different markers indicate the lowering of the magnetic field to match the decreasing temperature. 
In graph (b), the data points denoted by circles represent the decay of the number of atoms in the absence of cooling. 
The corresponding $1/e$ lifetime is $\tau=$\unit[27]{s}. 
(c) The PSD plotted versus the number of atoms allows us to determine the efficiency of the cooling from the slope of the resulting curve.}
\end{figure}
Figure \ref{fig:coolingplot}(a) shows the temperature as a function of time for an optimized cooling sequence, in which the magnetic field was held constant during the time intervals 0-5s, 5-7s, 7-11s and 11-16s at values of $\mathrm{B_1=\unit[350]{mG}}$, $\mathrm{B_2=\unit[230]{mG}}$, $\mathrm{B_3=\unit[160]{mG}}$ and $\mathrm{B_4=\unit[70]{mG}}$ respectively. 
We observe a decrease in temperature from $\mathrm{T_0=\unit[90]{\mu K}}$ to $\mathrm{T=\unit[15]{\mu K}}$ in the first 10 s and a further decrease to $\mathrm{T=\unit[11]{\mu K}}$ within the following 5 s. 
We achieve cooling rates more than 1 order of magnitude higher than those observed in Ref. \cite{fattori06}.  
We explain this improvement with the stronger confinement of our ODT and therefore the higher rate of dipolar relaxation collisions. 
The evolution of the number of atoms with and without cooling is presented in Fig. \ref{fig:coolingplot}(b). 
While the lifetime of the atoms in the absence of light is only limited by the background pressure, the demagnetization cooling is accompanied by additional losses of atoms, which were not observed in earlier work. 
We attribute these losses to excited-state collisions, possibly enhanced by re-absorption of scattered light in the dense medium. 
The adjustment of the magnetic field leads to an increase of dipolar relaxations and stronger losses, as discussed below. 
Nevertheless, the gain of the phase space density (PSD) over the loss in atom number, shown in Fig. \ref{fig:coolingplot}(c), is superior to evaporative cooling: the absolute value of the initial slope in the doubly logarithmic graph in Fig. \ref{fig:coolingplot}(c) amounts to 6 and exceeds typical values ($<4$) obtained in evaporative cooling. 
\begin{figure*}%
	\centering
	\includegraphics[scale=1]{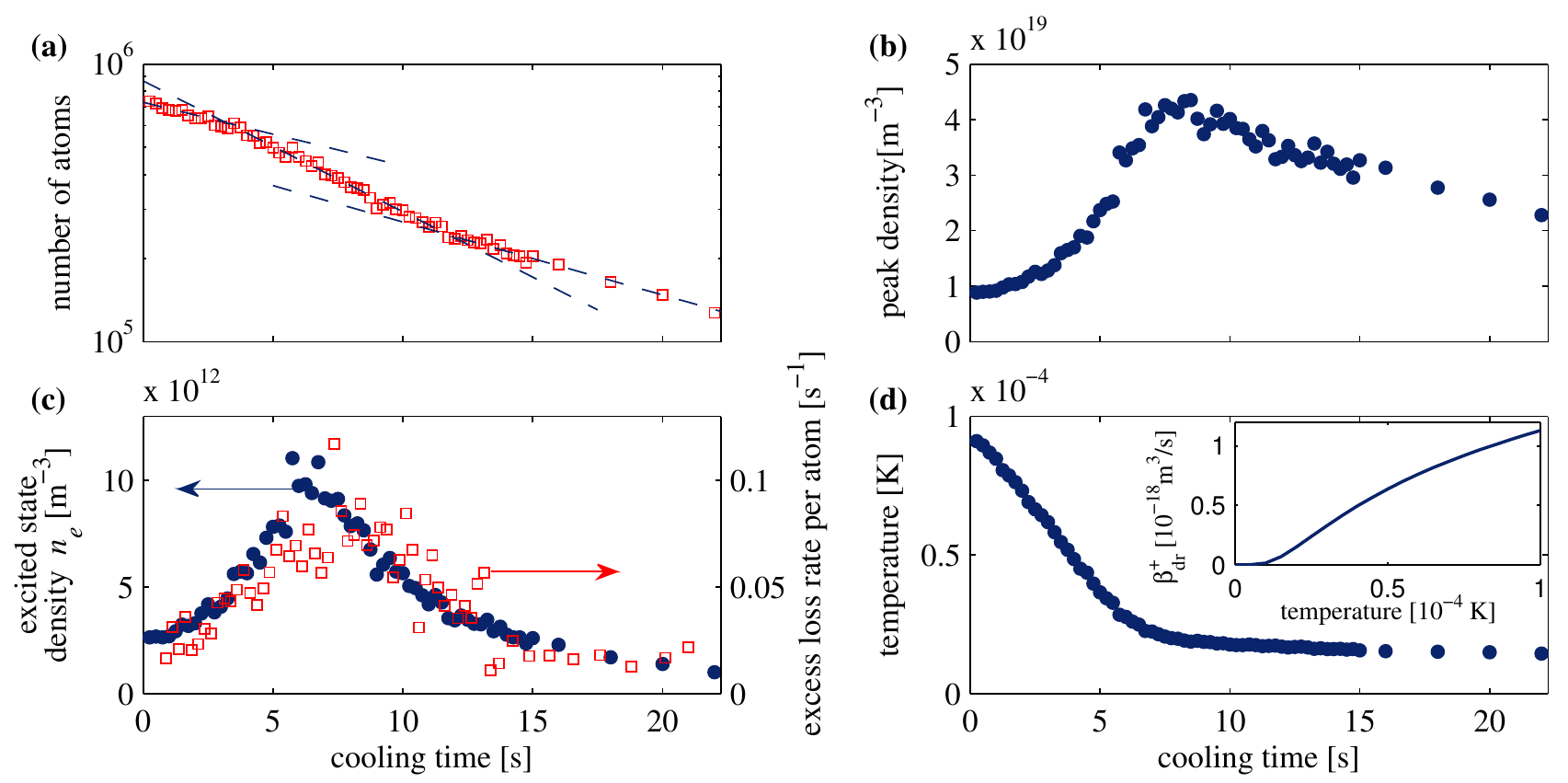}%
	\caption{\label{fig:rhosquare} Direct comparison of the increased loss of atoms and the calculated excited-state density. 
The evolution of the number of atoms in panel (a) is shown; the dashed lines are guides to the eye, emphasizing the changes of the decay rate. 
We use the measured atomic peak density (b) and the temperature-dependent dipolar relaxation coefficient $\beta_{dr}^+$ (d) to calculate the effective excited-state density $n_e$ [solid circles in panel (c)] according to Eq. (\ref{eq:ne2}). 
The empty squares in panel (c) are directly obtained from the change of number of atoms in panel (a) (see text) and correspond to the excess loss rate per atom.} 
\end{figure*}
To study the nature of the losses during the demagnetization cooling we performed additional experiments. 
In the case of excited-state collisions, an optical pumping photon is absorbed during the collision of two atoms. 
Conversion of internal excitation energy to kinetic energy as the atoms are accelerated by the interaction potential causes the loss of both atoms. 
Therefore, one expects these losses to scale with the density of the excited-state atoms and the density of the ground-state atoms. 
The number of excited-state atoms $N_e$ is proportional to the number of atoms that couple to the optical pumping light field $N_c$ (that is, all magnetic substates with $m_J>-3$) and the pumping rate $\Gamma_{\mathrm{OP}}$:
\begin{equation}
N_e=\frac{\Gamma_{\mathrm{OP}}}{\Gamma}N_c,
\label{eq:Ne}
\end{equation}
where $\Gamma=2\pi\times$\unit[5]{MHz} is the transition strength. 
Equation (\ref{eq:Ne}) is valid for $\Gamma_{\mathrm{OP}}\ll\Gamma$, which is the case for all our experiments.
We therefore obtain $N_e\ll N_c$. 
The dynamics of the populations of the magnetic substates are governed by a set of rate equations. 
Here, we first consider the case without losses and describe the system in a simplified picture with the dark-state population $N_{nc}$ in $m_J=-3$ and the coupled states population $N_c$, which is given by 
\begin{equation}
\dot{N}_c= \beta_{\mathrm{dr}}^+\frac{N_{nc}^2}{V} - \beta_{\mathrm{dr}}^-\frac{N_{nc}N_c}{V} - \Gamma_{\mathrm{OP}}N_c,
\label{eq:Nc}
\end{equation}
where $V$ is the temperature-dependent effective volume, $\beta_{\mathrm{dr}}^+$ is the dipolar relaxation coefficient which determines the rate of spin flips from the $m_J=-3$ state to higher states, $\beta_{\mathrm{dr}}^-$ is the coefficient in the reversed direction, and the last term describes optical pumping back to the dark state. 
The dipolar relaxation coefficients $\beta_{dr}^\pm$ depend on the temperature and the magnetic field \cite{hensler05}.
In order to estimate the general scaling of the populations we assume that at time scales of optical pumping and dipolar relaxations the temperature changes only slowly; therefore the volume and the coefficients $\beta_{\mathrm{dr}}^\pm$ are considered constant. 
As a consequence, dipolar relaxations in both directions and optical pumping are in equilibrium; hence 
\begin{equation}
\Gamma_{\mathrm{OP}}N_c + \beta_{\mathrm{dr}}^-\frac{N_{nc}N_c}{V} = \beta_{\mathrm{dr}}^+\frac{N_{nc}^2}{V}.
\label{eq:Nc0}
\end{equation}
Let us further consider two extreme cases. 
In the first case, the optical pumping is much faster than the back-relaxation: $\Gamma_{\mathrm{OP}}\gg\beta_{\mathrm{dr}}^-\frac{N_{nc}}{V}$.
In other words, atoms that are transferred to $m_J>-3$ by a dipolar relaxation, are immediately pumped back.
Then, we obtain
\begin{equation}
N_c \approx \beta_{\mathrm{dr}}^+\frac{N_{nc}^2}{\Gamma_{\mathrm{OP}}V}.
\label{eq:NcSat}
\end{equation}
In the opposite limit ($\Gamma_{\mathrm{OP}}\ll\beta_{\mathrm{dr}}^-\frac{N_{nc}}{V}$), the steady state is close to thermal equilibrium of the spin degree of freedom and we get
\begin{equation}
N_c\approx\frac{\beta_{\mathrm{dr}}^+}{\beta_{\mathrm{dr}}^-}N_{nc}.
\label{eq:NcSpin}
\end{equation}
Finally, let us consider the dynamics of the total number of ground-state atoms $N=N_{nc}+N_{c}$ including losses and given by
\begin{equation}
\dot{N} = - \gamma_{\mathrm{Bg}} N - \beta_{\mathrm{LC}}\frac{N N_e}{V}.
\label{eq:Ntot}
\end{equation}
The first term on the right-hand side corresponds to losses due to collisions with atoms from the background gas with the rate coefficient $\gamma_{\mathrm{Bg}}$, the second term accounts for light-assisted collisions with the rate coefficient $\beta_{\mathrm{LC}}$. 
Further we assume that $\beta_{\mathrm{LC}}$ is equal for all $m_J$ states.

In the first set of experiments, the calculated optical pumping rate was $\Gamma_{\mathrm{OP}}\approx\unit[350]{s^{-1}}$, which is much higher than the typical dipolar relaxation rates being on the order of $\unit[10]{s^{-1}}$.
We assume therefore that Eq. (\ref{eq:NcSat}) holds, yielding $N_{nc}\gg N_c$.
We plug Eq. (\ref{eq:Ne}) into Eq. (\ref{eq:Ntot}) and substitute $N_c$ with Eq. (\ref{eq:NcSat}). 
First, we find that the optical pumping rate $\Gamma_{\mathrm{OP}}$ drops out; thus we denote this regime as saturated.
Second, in this saturated regime the loss rate coefficient from excited-state collisions is proportional to the effective excited-state density given by 
\begin{equation}
n_e\equiv\frac{\beta_\mathrm{dr}^+}{\Gamma}\frac{N_{nc}^2}{V^2}\approx{}\frac{\beta_\mathrm{dr}^+}{\Gamma}\frac{N^2}{V^2}.
\label{eq:ne2}
\end{equation}
In order to observe this scaling we performed a cooling sequence at the constant magnetic field $\mathrm{B_1=\unit[350]{mG}}$.
We varied the cooling time and recorded the number of atoms, the peak density, and the temperature of the sample.
The results are plotted in Fig. \ref{fig:rhosquare}.
The number of atoms is plotted versus cooling time in Fig. \ref{fig:rhosquare}(a); the dashed lines emphasize the change of the loss rate over the time. 
In order to show the effect of light-assisted losses we compare the evolution of the loss rate per atom (denoted by empty squares) to the excited-state density $n_e$ (denoted by solid circles) in Fig. \ref{fig:rhosquare}(c).
We calculated $n_e$ for each data point  using the measured peak density (b) as $\frac{N}{V}$ and determining the dipolar relaxation coefficient $\beta_{\mathrm{dr}}^+$(inset) from the measured temperature in Fig. \ref{fig:rhosquare}(d). 
We obtained the loss rate by taking the derivative of the measured atom number evolution \footnote{The atom number data were smoothed by a moving average filter prior to performing the derivative operation.}; we then subtracted a background loss rate arising from the life time of \unit[27]{s}.
The resulting excess loss rate was then divided by the number of atoms, yielding, according to Eq. (\ref{eq:Ntot}) a quantity proportional to the excited state density: $(\dot{N}- \gamma_{\mathrm{Bg}} N)/N=\beta_{\mathrm{LC}}\frac{N_e}{V}$.
We find a strong temporal correlation between the variation of the measured rates of atom losses and the calculated excited state density based on our model given by Eq. (\ref{eq:ne2}).
This agreement indicates the validity of our model in the saturated regime.
Furthermore, data in Fig. \ref{fig:rhosquare} (c) allow us to estimate the light-assisted loss rate coefficient $\beta_{\mathrm{LC}}\approx\unit[8\cdot10^{-15}]{m^{3}s^{-1}}$, which is within 1 order of magnitude of the semiclassical limit \cite{julienne91}.
\begin{figure}
	\includegraphics[scale=1]{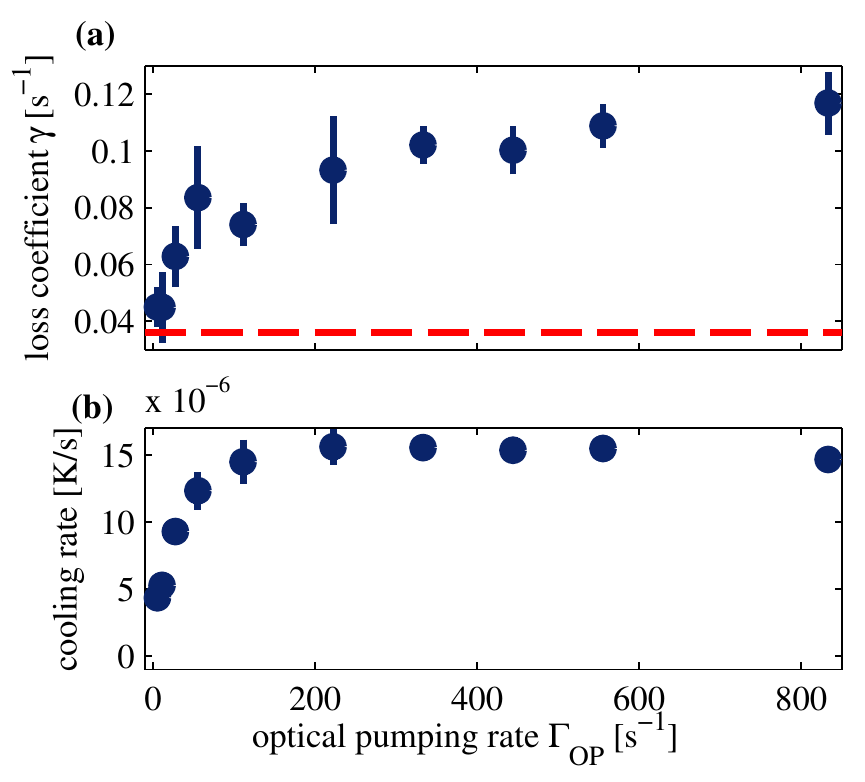}
	\caption{\label{fig:lossesVSpumpRF} Loss rate coefficient (a) and cooling rate (b) are plotted versus the optical pumping rate. 
	The dashed line in panel (a) corresponds to background losses in the absence of cooling. 
	The saturation of the cooling  and the losses indicate that the optical pumping rate exceeds the dipolar relaxation rate above $\unit[100]{s^{-1}}$. 
	The error bars correspond to confidence bounds of the fits.}
\end{figure}

In a second set of experiments we varied the intensity of the optical pumping light for each cooling sequence at the same magnetic field $\mathrm{B_1}$ as above.
We extracted the initial cooling rate from a linear fit to the first 4 s of the temperature evolution. 
Because of the complex interplay of  demagnetization cooling and light-assisted losses described above, we used an effective loss rate coefficient, $\gamma$, in order to compare different optical pumping light intensities.
We obtained $\gamma$ from an exponential fit to the number of atoms over the first 10 s.
Both the cooling rate and the loss rate coefficient are plotted as functions of the optical pumping rate in Fig. \ref{fig:lossesVSpumpRF}.
The data points corresponding to the lowest pumping light intensities are in the regime where $\Gamma_{\mathrm{OP}}\geq\beta_{\mathrm{dr}}^-\frac{N_{nc}}{V}$.
Therefore, substituting $N_c$ in Eq. (\ref{eq:Ne}) with Eq. (\ref{eq:NcSpin}) leads to a linear dependence of the excited state population on the optical pumping rate, similar to work on light-assisted collisions in magneto-optical traps\cite{Weiner99}.
As a consequence, for low optical pumping rates the cooling rate and the effective loss rate coefficient increase with $\Gamma_{\mathrm{OP}}$.
In the saturated regime, i.e.,  $\Gamma_{\mathrm{OP}}\gg\beta_{\mathrm{dr}}^-\frac{N_{nc}}{V}$, the cooling reaches a constant value.
The effective loss rate coefficients saturate (within error bars) for pumping rates above \unit[200]{$s^{-1}$}.
The residual increase of losses for larger intensities most likely originates from randomly polarized stray light, which is neglected in the rate equations.
A possible source of stray light is a pair of small coils \cite{falkenau11} within 1 mm of the atoms.

We investigated the decay of the number of atoms for different detunings in the accessible range of $\pm75\Gamma$. 
As expected for a constant optical pumping rate, we observed only minor variations of the total loss rate, with the exception of strong loss features close to the resonance. 
In contrast to previous work on detuning dependence \cite{Gallagher89}, we adjusted the intensity of the optical pumping light for each detuning setting to keep the optical pumping rate $\Gamma_{\mathrm{OP}}$ constant.
Experimental data presented in this work were taken at a detuning of $-72\Gamma$, at which we obtained the best cooling results.
However, for very large detunings on the order of \unit[100]{GHz}, excited-state collisions may be strongly suppressed when the internuclear distance of resonant absorption coincides with the node in the ground-state wave function \cite{Burnett96}. 
Attaining this regime requires significantly higher intensities and is the subject of future investigations.
\par 
From the presented arguments we conclude that collision and losses involving atoms in the excited state prevent us from reaching densities higher than $5\times 10^{19} m^{-3}$ using demagnetization cooling under given conditions. 
Eventually, the cooling stops when the relaxation rate stagnates because of a greatly reduced number of atoms. 
Although, we have observed temperatures down to $\mathrm{T=\unit[6]{\mu K}}$ and higher PSD by reducing the confinement of the dipole trap, ultimately, in order to achieve the onset of condensation at $\mathrm{T_{min}\approx T_R=\unit[1.02]{\mu K}}$, the density must exceed $\unit[10^{20}]{m^{-3}}$. Due to above mentioned excited state losses this cannot be reached within the present experimental parameters.

\par

In the following, we are going to elaborate why the possibility of achieving Bose-Einstein condensation by means of demagnetization cooling is more favourable using dysprosium instead of chromium. 
First, the cooling works faster since the cross section for dipolar relaxation  scales with mass $m$ and electronic angular momentum $J$ as follows \cite{hensler03}:
\begin{equation}
	\sigma\propto \frac{8\pi}{15}J^3\left(\frac{\mu_0 (g_J\mu_B)^2 m}{4\pi\hbar} \right)^2.
\end{equation}
Compared to $^{52}$Cr,  $^{164}$Dy has three times higher mass and higher total electronic angular momentum ($J=3$ for $^{52}$Cr and $J=8$ for $^{164}$Dy), whereas the Land\'{e}-factor $g_J$ is larger in $^{52}$Cr ($g_{\mathrm{Cr}}/g_{\mathrm{Dy}}=1.65$). 
Therefore, the resulting cross section is 22 times larger for $^{164}$Dy, which would allow working at a higher Zeeman splitting compared to the temperature, resulting in higher energy dissipation per photon.

The recoil temperature for $^{164}$Dy is lower  than for $^{52}$Cr due to its higher mass and the larger wavelength of the pumping transition.
As a consequence, the critical density at $\mathrm{T_{min}\approx T_R}$ is reduced by a factor of $(\frac{\lambda_{\mathrm{Cr}}}{\lambda_{\mathrm{Dy}}})^2=4.11$. 
This scaling results in a critical density below the maximal density observed in our experiment. Therefore we expect that  demagnetization cooling to degeneracy should be possible with dysprosium.
\par

In conclusion, we have shown that demagnetization cooling can be used over a wide temperature range, provided dipolar relaxation ensures thermal equilibrium between the motional and the spin degrees of freedom. The high efficiency of the demagnetization cooling stems from the fact that each scattered photon dissipates energy on the order of $k_BT$, which is much larger than in usual laser cooling techniques. 
It allows reaching a density up to $5\times 10^{19} m^{-3}$ with a PSD$\approx0.02$. 
In our experiment the number of scattered photons is on the order of 10 photons per atom. 
It becomes especially important, when a pumping transition is not closed and only a limited number of photons can be scattered.
We have demonstrated that the limiting mechanism is the loss of atoms resulting from collisions between excited- and ground-state atoms. 
This limiting factor has not yet been considered in the context of demagnetization cooling and shows an unusually strong scaling with the density of the atoms in the ground-state. 
Light-assisted losses turn out to be a serious obstacle for $^{52}$Cr to reach degeneracy by means of demagnetization cooling only.

This work was supported by the DFG under Contract No. PF381/14-1. V.V.V. acknowledges support from the Friedrich-Ebert-Stiftung.


\begin{thebibliography}{28}%
\makeatletter
\providecommand \@ifxundefined [1]{%
 \@ifx{#1\undefined}
}%
\providecommand \@ifnum [1]{%
 \ifnum #1\expandafter \@firstoftwo
 \else \expandafter \@secondoftwo
 \fi
}%
\providecommand \@ifx [1]{%
 \ifx #1\expandafter \@firstoftwo
 \else \expandafter \@secondoftwo
 \fi
}%
\providecommand \natexlab [1]{#1}%
\providecommand \enquote  [1]{``#1''}%
\providecommand \bibnamefont  [1]{#1}%
\providecommand \bibfnamefont [1]{#1}%
\providecommand \citenamefont [1]{#1}%
\providecommand \href@noop [0]{\@secondoftwo}%
\providecommand \href [0]{\begingroup \@sanitize@url \@href}%
\providecommand \@href[1]{\@@startlink{#1}\@@href}%
\providecommand \@@href[1]{\endgroup#1\@@endlink}%
\providecommand \@sanitize@url [0]{\catcode `\\12\catcode `\$12\catcode
  `\&12\catcode `\#12\catcode `\^12\catcode `\_12\catcode `\%12\relax}%
\providecommand \@@startlink[1]{}%
\providecommand \@@endlink[0]{}%
\providecommand \url  [0]{\begingroup\@sanitize@url \@url }%
\providecommand \@url [1]{\endgroup\@href {#1}{\urlprefix }}%
\providecommand \urlprefix  [0]{URL }%
\providecommand \Eprint [0]{\href }%
\providecommand \doibase [0]{http://dx.doi.org/}%
\providecommand \selectlanguage [0]{\@gobble}%
\providecommand \bibinfo  [0]{\@secondoftwo}%
\providecommand \bibfield  [0]{\@secondoftwo}%
\providecommand \translation [1]{[#1]}%
\providecommand \BibitemOpen [0]{}%
\providecommand \bibitemStop [0]{}%
\providecommand \bibitemNoStop [0]{.\EOS\space}%
\providecommand \EOS [0]{\spacefactor3000\relax}%
\providecommand \BibitemShut  [1]{\csname bibitem#1\endcsname}%
\let\auto@bib@innerbib\@empty
\bibitem [{\citenamefont {Lu}\ \emph {et~al.}(2011)\citenamefont {Lu},
  \citenamefont {Burdick}, \citenamefont {Youn},\ and\ \citenamefont
  {Lev}}]{Lu11b}%
  \BibitemOpen
  \bibfield  {author} {\bibinfo {author} {\bibfnamefont {M.}~\bibnamefont
  {Lu}}, \bibinfo {author} {\bibfnamefont {N.~Q.}\ \bibnamefont {Burdick}},
  \bibinfo {author} {\bibfnamefont {S.~H.}\ \bibnamefont {Youn}}, \ and\
  \bibinfo {author} {\bibfnamefont {B.~L.}\ \bibnamefont {Lev}},\ }\href
  {\doibase 10.1103/PhysRevLett.107.190401} {\bibfield  {journal} {\bibinfo
  {journal} {Phys. Rev. Lett.}\ }\textbf {\bibinfo {volume} {107}},\ \bibinfo
  {pages} {190401} (\bibinfo {year} {2011})}\BibitemShut {NoStop}%
\bibitem [{\citenamefont {Aikawa}\ \emph {et~al.}(2012)\citenamefont {Aikawa},
  \citenamefont {Frisch}, \citenamefont {Mark}, \citenamefont {Baier},
  \citenamefont {Rietzler}, \citenamefont {Grimm},\ and\ \citenamefont
  {Ferlaino}}]{Aikawa12}%
  \BibitemOpen
  \bibfield  {author} {\bibinfo {author} {\bibfnamefont {K.}~\bibnamefont
  {Aikawa}}, \bibinfo {author} {\bibfnamefont {A.}~\bibnamefont {Frisch}},
  \bibinfo {author} {\bibfnamefont {M.}~\bibnamefont {Mark}}, \bibinfo {author}
  {\bibfnamefont {S.}~\bibnamefont {Baier}}, \bibinfo {author} {\bibfnamefont
  {A.}~\bibnamefont {Rietzler}}, \bibinfo {author} {\bibfnamefont
  {R.}~\bibnamefont {Grimm}}, \ and\ \bibinfo {author} {\bibfnamefont
  {F.}~\bibnamefont {Ferlaino}},\ }\href {\doibase
  10.1103/PhysRevLett.108.210401} {\bibfield  {journal} {\bibinfo  {journal}
  {Phys. Rev. Lett.}\ }\textbf {\bibinfo {volume} {108}},\ \bibinfo {pages}
  {210401} (\bibinfo {year} {2012})}\BibitemShut {NoStop}%
\bibitem [{\citenamefont {Stellmer}\ \emph {et~al.}(2009)\citenamefont
  {Stellmer}, \citenamefont {Tey}, \citenamefont {Huang}, \citenamefont
  {Grimm},\ and\ \citenamefont {Schreck}}]{Stellmer09}%
  \BibitemOpen
  \bibfield  {author} {\bibinfo {author} {\bibfnamefont {S.}~\bibnamefont
  {Stellmer}}, \bibinfo {author} {\bibfnamefont {M.~K.}\ \bibnamefont {Tey}},
  \bibinfo {author} {\bibfnamefont {B.}~\bibnamefont {Huang}}, \bibinfo
  {author} {\bibfnamefont {R.}~\bibnamefont {Grimm}}, \ and\ \bibinfo {author}
  {\bibfnamefont {F.}~\bibnamefont {Schreck}},\ }\href {\doibase
  10.1103/PhysRevLett.103.200401} {\bibfield  {journal} {\bibinfo  {journal}
  {Phys. Rev. Lett.}\ }\textbf {\bibinfo {volume} {103}},\ \bibinfo {pages}
  {200401} (\bibinfo {year} {2009})}\BibitemShut {NoStop}%
\bibitem [{\citenamefont {Kraft}\ \emph {et~al.}(2009)\citenamefont {Kraft},
  \citenamefont {Vogt}, \citenamefont {Appel}, \citenamefont {Riehle},\ and\
  \citenamefont {Sterr}}]{Kraft09}%
  \BibitemOpen
  \bibfield  {author} {\bibinfo {author} {\bibfnamefont {S.}~\bibnamefont
  {Kraft}}, \bibinfo {author} {\bibfnamefont {F.}~\bibnamefont {Vogt}},
  \bibinfo {author} {\bibfnamefont {O.}~\bibnamefont {Appel}}, \bibinfo
  {author} {\bibfnamefont {F.}~\bibnamefont {Riehle}}, \ and\ \bibinfo {author}
  {\bibfnamefont {U.}~\bibnamefont {Sterr}},\ }\href {\doibase
  10.1103/PhysRevLett.103.130401} {\bibfield  {journal} {\bibinfo  {journal}
  {Phys. Rev. Lett.}\ }\textbf {\bibinfo {volume} {103}},\ \bibinfo {pages}
  {130401} (\bibinfo {year} {2009})}\BibitemShut {NoStop}%
\bibitem [{\citenamefont {Takasu}\ \emph {et~al.}(2003)\citenamefont {Takasu},
  \citenamefont {Maki}, \citenamefont {Komori}, \citenamefont {Takano},
  \citenamefont {Honda}, \citenamefont {Kumakura}, \citenamefont {Yabuzaki},\
  and\ \citenamefont {Takahashi}}]{Takasu03}%
  \BibitemOpen
  \bibfield  {author} {\bibinfo {author} {\bibfnamefont {Y.}~\bibnamefont
  {Takasu}}, \bibinfo {author} {\bibfnamefont {K.}~\bibnamefont {Maki}},
  \bibinfo {author} {\bibfnamefont {K.}~\bibnamefont {Komori}}, \bibinfo
  {author} {\bibfnamefont {T.}~\bibnamefont {Takano}}, \bibinfo {author}
  {\bibfnamefont {K.}~\bibnamefont {Honda}}, \bibinfo {author} {\bibfnamefont
  {M.}~\bibnamefont {Kumakura}}, \bibinfo {author} {\bibfnamefont
  {T.}~\bibnamefont {Yabuzaki}}, \ and\ \bibinfo {author} {\bibfnamefont
  {Y.}~\bibnamefont {Takahashi}},\ }\href {\doibase
  10.1103/PhysRevLett.91.040404} {\bibfield  {journal} {\bibinfo  {journal}
  {Phys. Rev. Lett.}\ }\textbf {\bibinfo {volume} {91}},\ \bibinfo {pages}
  {040404} (\bibinfo {year} {2003})}\BibitemShut {NoStop}%
\bibitem [{\citenamefont {Griesmaier}\ \emph {et~al.}(2005)\citenamefont
  {Griesmaier}, \citenamefont {Werner}, \citenamefont {Hensler}, \citenamefont
  {Stuhler},\ and\ \citenamefont {Pfau}}]{griesmaier05}%
  \BibitemOpen
  \bibfield  {author} {\bibinfo {author} {\bibfnamefont {A.}~\bibnamefont
  {Griesmaier}}, \bibinfo {author} {\bibfnamefont {J.}~\bibnamefont {Werner}},
  \bibinfo {author} {\bibfnamefont {S.}~\bibnamefont {Hensler}}, \bibinfo
  {author} {\bibfnamefont {J.}~\bibnamefont {Stuhler}}, \ and\ \bibinfo
  {author} {\bibfnamefont {T.}~\bibnamefont {Pfau}},\ }\href {\doibase
  10.1103/PhysRevLett.94.160401} {\bibfield  {journal} {\bibinfo  {journal}
  {Phys. Rev. Lett.}\ }\textbf {\bibinfo {volume} {94}},\ \bibinfo {pages}
  {160401} (\bibinfo {year} {2005})}\BibitemShut {NoStop}%
\bibitem [{\citenamefont {G\'oral}\ \emph {et~al.}(2002)\citenamefont
  {G\'oral}, \citenamefont {Santos},\ and\ \citenamefont
  {Lewenstein}}]{Goral02}%
  \BibitemOpen
  \bibfield  {author} {\bibinfo {author} {\bibfnamefont {K.}~\bibnamefont
  {G\'oral}}, \bibinfo {author} {\bibfnamefont {L.}~\bibnamefont {Santos}}, \
  and\ \bibinfo {author} {\bibfnamefont {M.}~\bibnamefont {Lewenstein}},\
  }\href {\doibase 10.1103/PhysRevLett.88.170406} {\bibfield  {journal}
  {\bibinfo  {journal} {Phys. Rev. Lett.}\ }\textbf {\bibinfo {volume} {88}},\
  \bibinfo {pages} {170406} (\bibinfo {year} {2002})}\BibitemShut {NoStop}%
\bibitem [{\citenamefont {Pedri}\ and\ \citenamefont {Santos}(2005)}]{Pedri05}%
  \BibitemOpen
  \bibfield  {author} {\bibinfo {author} {\bibfnamefont {P.}~\bibnamefont
  {Pedri}}\ and\ \bibinfo {author} {\bibfnamefont {L.}~\bibnamefont {Santos}},\
  }\href {\doibase 10.1103/PhysRevLett.95.200404} {\bibfield  {journal}
  {\bibinfo  {journal} {Phys. Rev. Lett.}\ }\textbf {\bibinfo {volume} {95}},\
  \bibinfo {pages} {200404} (\bibinfo {year} {2005})}\BibitemShut {NoStop}%
\bibitem [{\citenamefont {Santos}\ \emph {et~al.}(2003)\citenamefont {Santos},
  \citenamefont {Shlyapnikov},\ and\ \citenamefont {Lewenstein}}]{Santos03}%
  \BibitemOpen
  \bibfield  {author} {\bibinfo {author} {\bibfnamefont {L.}~\bibnamefont
  {Santos}}, \bibinfo {author} {\bibfnamefont {G.~V.}\ \bibnamefont
  {Shlyapnikov}}, \ and\ \bibinfo {author} {\bibfnamefont {M.}~\bibnamefont
  {Lewenstein}},\ }\href {\doibase 10.1103/PhysRevLett.90.250403} {\bibfield
  {journal} {\bibinfo  {journal} {Phys. Rev. Lett.}\ }\textbf {\bibinfo
  {volume} {90}},\ \bibinfo {pages} {250403} (\bibinfo {year}
  {2003})}\BibitemShut {NoStop}%
\bibitem [{\citenamefont {Katori}\ \emph {et~al.}(1999)\citenamefont {Katori},
  \citenamefont {Ido}, \citenamefont {Isoya},\ and\ \citenamefont
  {Kuwata-Gonokami}}]{Katori99}%
  \BibitemOpen
  \bibfield  {author} {\bibinfo {author} {\bibfnamefont {H.}~\bibnamefont
  {Katori}}, \bibinfo {author} {\bibfnamefont {T.}~\bibnamefont {Ido}},
  \bibinfo {author} {\bibfnamefont {Y.}~\bibnamefont {Isoya}}, \ and\ \bibinfo
  {author} {\bibfnamefont {M.}~\bibnamefont {Kuwata-Gonokami}},\ }\href
  {\doibase 10.1103/PhysRevLett.82.1116} {\bibfield  {journal} {\bibinfo
  {journal} {Phys. Rev. Lett.}\ }\textbf {\bibinfo {volume} {82}},\ \bibinfo
  {pages} {1116} (\bibinfo {year} {1999})}\BibitemShut {NoStop}%
\bibitem [{\citenamefont {Frisch}\ \emph {et~al.}(2012)\citenamefont {Frisch},
  \citenamefont {Aikawa}, \citenamefont {Mark}, \citenamefont {Rietzler},
  \citenamefont {Schindler}, \citenamefont {Zupani\ifmmode~\check{c}\else
  \v{c}\fi{}}, \citenamefont {Grimm},\ and\ \citenamefont
  {Ferlaino}}]{Frisch12}%
  \BibitemOpen
  \bibfield  {author} {\bibinfo {author} {\bibfnamefont {A.}~\bibnamefont
  {Frisch}}, \bibinfo {author} {\bibfnamefont {K.}~\bibnamefont {Aikawa}},
  \bibinfo {author} {\bibfnamefont {M.}~\bibnamefont {Mark}}, \bibinfo {author}
  {\bibfnamefont {A.}~\bibnamefont {Rietzler}}, \bibinfo {author}
  {\bibfnamefont {J.}~\bibnamefont {Schindler}}, \bibinfo {author}
  {\bibfnamefont {E.}~\bibnamefont {Zupani\ifmmode~\check{c}\else \v{c}\fi{}}},
  \bibinfo {author} {\bibfnamefont {R.}~\bibnamefont {Grimm}}, \ and\ \bibinfo
  {author} {\bibfnamefont {F.}~\bibnamefont {Ferlaino}},\ }\href {\doibase
  10.1103/PhysRevA.85.051401} {\bibfield  {journal} {\bibinfo  {journal} {Phys.
  Rev. A}\ }\textbf {\bibinfo {volume} {85}},\ \bibinfo {pages} {051401}
  (\bibinfo {year} {2012})}\BibitemShut {NoStop}%
\bibitem [{\citenamefont {Zeppenfeld}\ \emph {et~al.}(2012)\citenamefont
  {Zeppenfeld}, \citenamefont {Englert}, \citenamefont {Gl\"{o}öckner},
  \citenamefont {Prehn}, \citenamefont {Mielenz}, \citenamefont {Sommer},
  \citenamefont {van Buuren}, \citenamefont {Motsch},\ and\ \citenamefont
  {Rempe}}]{zeppenfeld12}%
  \BibitemOpen
  \bibfield  {author} {\bibinfo {author} {\bibfnamefont {M.}~\bibnamefont
  {Zeppenfeld}}, \bibinfo {author} {\bibfnamefont {B.~G.~U.}\ \bibnamefont
  {Englert}}, \bibinfo {author} {\bibfnamefont {R.}~\bibnamefont
  {Gl\"{o}öckner}}, \bibinfo {author} {\bibfnamefont {A.}~\bibnamefont
  {Prehn}}, \bibinfo {author} {\bibfnamefont {M.}~\bibnamefont {Mielenz}},
  \bibinfo {author} {\bibfnamefont {C.}~\bibnamefont {Sommer}}, \bibinfo
  {author} {\bibfnamefont {L.~D.}\ \bibnamefont {van Buuren}}, \bibinfo
  {author} {\bibfnamefont {M.}~\bibnamefont {Motsch}}, \ and\ \bibinfo {author}
  {\bibfnamefont {G.}~\bibnamefont {Rempe}},\ }\href {\doibase
  10.1038/nature11595} {\bibfield  {journal} {\bibinfo  {journal} {Nature}\
  }\textbf {\bibinfo {volume} {491}},\ \bibinfo {pages} {570} (\bibinfo {year}
  {2012})}\BibitemShut {NoStop}%
\bibitem [{\citenamefont {Shuman}\ \emph {et~al.}(2010)\citenamefont {Shuman},
  \citenamefont {Barry},\ and\ \citenamefont {DeMille}}]{shuman10}%
  \BibitemOpen
  \bibfield  {author} {\bibinfo {author} {\bibfnamefont {E.~S.}\ \bibnamefont
  {Shuman}}, \bibinfo {author} {\bibfnamefont {J.~F.}\ \bibnamefont {Barry}}, \
  and\ \bibinfo {author} {\bibfnamefont {D.}~\bibnamefont {DeMille}},\
  }\href@noop {} {\bibfield  {journal} {\bibinfo  {journal} {Nature (London)}\
  ,\ \bibinfo {pages} {820}} (\bibinfo {year} {2010})}\BibitemShut {NoStop}%
\bibitem [{\citenamefont {McClelland}\ and\ \citenamefont
  {Hanssen}(2006)}]{mcclelland06}%
  \BibitemOpen
  \bibfield  {author} {\bibinfo {author} {\bibfnamefont {J.~J.}\ \bibnamefont
  {McClelland}}\ and\ \bibinfo {author} {\bibfnamefont {J.~L.}\ \bibnamefont
  {Hanssen}},\ }\href {\doibase 10.1103/PhysRevLett.96.143005} {\bibfield
  {journal} {\bibinfo  {journal} {Phys. Rev. Lett.}\ }\textbf {\bibinfo
  {volume} {96}},\ \bibinfo {pages} {143005} (\bibinfo {year}
  {2006})}\BibitemShut {NoStop}%
\bibitem [{\citenamefont {Cirac}\ and\ \citenamefont
  {Lewenstein}(1996)}]{BAR96}%
  \BibitemOpen
  \bibfield  {author} {\bibinfo {author} {\bibfnamefont {J.~I.}\ \bibnamefont
  {Cirac}}\ and\ \bibinfo {author} {\bibfnamefont {M.}~\bibnamefont
  {Lewenstein}},\ }\href {\doibase 10.1103/PhysRevA.53.2466} {\bibfield
  {journal} {\bibinfo  {journal} {Phys. Rev. A}\ }\textbf {\bibinfo {volume}
  {53}},\ \bibinfo {pages} {2466} (\bibinfo {year} {1996})}\BibitemShut
  {NoStop}%
\bibitem [{\citenamefont {Santos}\ \emph {et~al.}(2001)\citenamefont {Santos},
  \citenamefont {Floegel}, \citenamefont {Pfau},\ and\ \citenamefont
  {Lewenstein}}]{santos01}%
  \BibitemOpen
  \bibfield  {author} {\bibinfo {author} {\bibfnamefont {L.}~\bibnamefont
  {Santos}}, \bibinfo {author} {\bibfnamefont {F.}~\bibnamefont {Floegel}},
  \bibinfo {author} {\bibfnamefont {T.}~\bibnamefont {Pfau}}, \ and\ \bibinfo
  {author} {\bibfnamefont {M.}~\bibnamefont {Lewenstein}},\ }\href {\doibase
  10.1103/PhysRevA.63.063408} {\bibfield  {journal} {\bibinfo  {journal} {Phys.
  Rev. A}\ }\textbf {\bibinfo {volume} {63}},\ \bibinfo {pages} {063408}
  (\bibinfo {year} {2001})}\BibitemShut {NoStop}%
\bibitem [{\citenamefont {Stellmer}\ \emph {et~al.}(2013)\citenamefont
  {Stellmer}, \citenamefont {Pasquiou}, \citenamefont {Grimm},\ and\
  \citenamefont {Schreck}}]{Stellmer13}%
  \BibitemOpen
  \bibfield  {author} {\bibinfo {author} {\bibfnamefont {S.}~\bibnamefont
  {Stellmer}}, \bibinfo {author} {\bibfnamefont {B.}~\bibnamefont {Pasquiou}},
  \bibinfo {author} {\bibfnamefont {R.}~\bibnamefont {Grimm}}, \ and\ \bibinfo
  {author} {\bibfnamefont {F.}~\bibnamefont {Schreck}},\ }\href {\doibase
  10.1103/PhysRevLett.110.263003} {\bibfield  {journal} {\bibinfo  {journal}
  {Phys. Rev. Lett.}\ }\textbf {\bibinfo {volume} {110}},\ \bibinfo {pages}
  {263003} (\bibinfo {year} {2013})}\BibitemShut {NoStop}%
\bibitem [{\citenamefont {Kastler}(1950)}]{kastler50}%
  \BibitemOpen
  \bibfield  {author} {\bibinfo {author} {\bibfnamefont {A.}~\bibnamefont
  {Kastler}},\ }\href {\doibase 10.1051/jphysrad:01950001106025500} {\bibfield
  {journal} {\bibinfo  {journal} {Le Journal de Physique et le Radium}\
  }\textbf {\bibinfo {volume} {11}},\ \bibinfo {pages} {255} (\bibinfo {year}
  {1950})}\BibitemShut {NoStop}%
\bibitem [{\citenamefont {Fattori}\ \emph {et~al.}(2006)\citenamefont
  {Fattori}, \citenamefont {Koch}, \citenamefont {Goetz}, \citenamefont
  {Griesmaier}, \citenamefont {Hensler}, \citenamefont {Stuhler},\ and\
  \citenamefont {Pfau}}]{fattori06}%
  \BibitemOpen
  \bibfield  {author} {\bibinfo {author} {\bibfnamefont {M.}~\bibnamefont
  {Fattori}}, \bibinfo {author} {\bibfnamefont {T.}~\bibnamefont {Koch}},
  \bibinfo {author} {\bibfnamefont {S.}~\bibnamefont {Goetz}}, \bibinfo
  {author} {\bibfnamefont {A.}~\bibnamefont {Griesmaier}}, \bibinfo {author}
  {\bibfnamefont {S.}~\bibnamefont {Hensler}}, \bibinfo {author} {\bibfnamefont
  {J.}~\bibnamefont {Stuhler}}, \ and\ \bibinfo {author} {\bibfnamefont
  {T.}~\bibnamefont {Pfau}},\ }\href@noop {} {\bibfield  {journal} {\bibinfo
  {journal} {Nature Physics}\ }\textbf {\bibinfo {volume} {2}} (\bibinfo {year}
  {2006})}\BibitemShut {NoStop}%
\bibitem [{\citenamefont {Hensler}\ \emph {et~al.}(2005)\citenamefont
  {Hensler}, \citenamefont {Greiner}, \citenamefont {Stuhler},\ and\
  \citenamefont {Pfau}}]{hensler05}%
  \BibitemOpen
  \bibfield  {author} {\bibinfo {author} {\bibfnamefont {S.}~\bibnamefont
  {Hensler}}, \bibinfo {author} {\bibfnamefont {A.}~\bibnamefont {Greiner}},
  \bibinfo {author} {\bibfnamefont {J.}~\bibnamefont {Stuhler}}, \ and\
  \bibinfo {author} {\bibfnamefont {T.}~\bibnamefont {Pfau}},\ }\href
  {http://stacks.iop.org/0295-5075/71/i=6/a=918} {\bibfield  {journal}
  {\bibinfo  {journal} {EPL (Europhysics Letters)}\ }\textbf {\bibinfo {volume}
  {71}},\ \bibinfo {pages} {918} (\bibinfo {year} {2005})}\BibitemShut
  {NoStop}%
\bibitem [{\citenamefont {Falkenau}\ \emph {et~al.}(2011)\citenamefont
  {Falkenau}, \citenamefont {Volchkov}, \citenamefont {R\"uhrig}, \citenamefont
  {Griesmaier},\ and\ \citenamefont {Pfau}}]{falkenau11}%
  \BibitemOpen
  \bibfield  {author} {\bibinfo {author} {\bibfnamefont {M.}~\bibnamefont
  {Falkenau}}, \bibinfo {author} {\bibfnamefont {V.~V.}\ \bibnamefont
  {Volchkov}}, \bibinfo {author} {\bibfnamefont {J.}~\bibnamefont {R\"uhrig}},
  \bibinfo {author} {\bibfnamefont {A.}~\bibnamefont {Griesmaier}}, \ and\
  \bibinfo {author} {\bibfnamefont {T.}~\bibnamefont {Pfau}},\ }\href {\doibase
  10.1103/PhysRevLett.106.163002} {\bibfield  {journal} {\bibinfo  {journal}
  {Phys. Rev. Lett.}\ }\textbf {\bibinfo {volume} {106}},\ \bibinfo {pages}
  {163002} (\bibinfo {year} {2011})}\BibitemShut {NoStop}%
\bibitem [{\citenamefont {Volchkov}\ \emph {et~al.}(2013)\citenamefont
  {Volchkov}, \citenamefont {R\"uhrig}, \citenamefont {Pfau},\ and\ \citenamefont
  {Griesmaier}}]{volchkov13}%
  \BibitemOpen
  \bibfield  {author} {\bibinfo {author} {\bibfnamefont {V.~V.}\ \bibnamefont
  {Volchkov}}, \bibinfo {author} {\bibfnamefont {J.}~\bibnamefont {R\"uhrig}},
  \bibinfo {author} {\bibfnamefont {T.}~\bibnamefont {Pfau}}, \ and\ \bibinfo
  {author} {\bibfnamefont {A.}~\bibnamefont {Griesmaier}},\ }\href
  {http://stacks.iop.org/1367-2630/15/i=9/a=093012} {\bibfield  {journal}
  {\bibinfo  {journal} {New Journal of Physics}\ }\textbf {\bibinfo {volume}
  {15}},\ \bibinfo {pages} {093012} (\bibinfo {year} {2013})}\BibitemShut
  {NoStop}%
\bibitem [{Note1()}]{Note1}%
  \BibitemOpen
  \bibinfo {note} {The atom number data were smoothed by a moving average
  filter prior to performing the derivative operation.}\BibitemShut {Stop}%
\bibitem [{\citenamefont {Julienne}\ and\ \citenamefont
  {Vigu\'e}(1991)}]{julienne91}%
  \BibitemOpen
  \bibfield  {author} {\bibinfo {author} {\bibfnamefont {P.~S.}\ \bibnamefont
  {Julienne}}\ and\ \bibinfo {author} {\bibfnamefont {J.}~\bibnamefont
  {Vigu\'e}},\ }\href {\doibase 10.1103/PhysRevA.44.4464} {\bibfield  {journal}
  {\bibinfo  {journal} {Phys. Rev. A}\ }\textbf {\bibinfo {volume} {44}},\
  \bibinfo {pages} {4464} (\bibinfo {year} {1991})}\BibitemShut {NoStop}%
\bibitem [{\citenamefont {Weiner}\ \emph {et~al.}(1999)\citenamefont {Weiner},
  \citenamefont {Bagnato}, \citenamefont {Zilio},\ and\ \citenamefont
  {Julienne}}]{Weiner99}%
  \BibitemOpen
  \bibfield  {author} {\bibinfo {author} {\bibfnamefont {J.}~\bibnamefont
  {Weiner}}, \bibinfo {author} {\bibfnamefont {V.~S.}\ \bibnamefont {Bagnato}},
  \bibinfo {author} {\bibfnamefont {S.}~\bibnamefont {Zilio}}, \ and\ \bibinfo
  {author} {\bibfnamefont {P.~S.}\ \bibnamefont {Julienne}},\ }\href {\doibase
  10.1103/RevModPhys.71.1} {\bibfield  {journal} {\bibinfo  {journal} {Rev.
  Mod. Phys.}\ }\textbf {\bibinfo {volume} {71}},\ \bibinfo {pages} {1}
  (\bibinfo {year} {1999})}\BibitemShut {NoStop}%
\bibitem [{\citenamefont {Gallagher}\ and\ \citenamefont
  {Pritchard}(1989)}]{Gallagher89}%
  \BibitemOpen
  \bibfield  {author} {\bibinfo {author} {\bibfnamefont {A.}~\bibnamefont
  {Gallagher}}\ and\ \bibinfo {author} {\bibfnamefont {D.~E.}\ \bibnamefont
  {Pritchard}},\ }\href {\doibase 10.1103/PhysRevLett.63.957} {\bibfield
  {journal} {\bibinfo  {journal} {Phys. Rev. Lett.}\ }\textbf {\bibinfo
  {volume} {63}},\ \bibinfo {pages} {957} (\bibinfo {year} {1989})}\BibitemShut
  {NoStop}%
\bibitem [{\citenamefont {Burnett}\ \emph {et~al.}(1996)\citenamefont
  {Burnett}, \citenamefont {Julienne},\ and\ \citenamefont
  {Suominen}}]{Burnett96}%
  \BibitemOpen
  \bibfield  {author} {\bibinfo {author} {\bibfnamefont {K.}~\bibnamefont
  {Burnett}}, \bibinfo {author} {\bibfnamefont {P.~S.}\ \bibnamefont
  {Julienne}}, \ and\ \bibinfo {author} {\bibfnamefont {K.-A.}\ \bibnamefont
  {Suominen}},\ }\href {\doibase 10.1103/PhysRevLett.77.1416} {\bibfield
  {journal} {\bibinfo  {journal} {Phys. Rev. Lett.}\ }\textbf {\bibinfo
  {volume} {77}},\ \bibinfo {pages} {1416} (\bibinfo {year}
  {1996})}\BibitemShut {NoStop}%
\bibitem [{\citenamefont {Hensler}\ \emph {et~al.}(2003)\citenamefont
  {Hensler}, \citenamefont {A.G\"{o}rlitz}, \citenamefont {S.Giovanazzi},\ and\
  \citenamefont {T.Pfau}}]{hensler03}%
  \BibitemOpen
  \bibfield  {author} {\bibinfo {author} {\bibfnamefont {S.}~\bibnamefont
  {Hensler}}, \bibinfo {author} {\bibnamefont {A.G\"{o}rlitz}}, \bibinfo
  {author} {\bibnamefont {S.Giovanazzi}}, \ and\ \bibinfo {author}
  {\bibnamefont {T.Pfau}},\ }\href@noop {} {\bibfield  {journal} {\bibinfo
  {journal} {Appl.Phys.B}\ }\textbf {\bibinfo {volume} {77}},\ \bibinfo {pages}
  {765} (\bibinfo {year} {2003})}\BibitemShut {NoStop}%
\end{thebibliography}
\end{document}